\input epsf
\documentstyle[prc,aps,twocolumn]{revtex}
\newcommand{\mmoy}{\langle m \rangle}
\newcommand{\Zmoy}{\langle Z \rangle}

\newcommand{\ztot}{z_{\rm tot}}
\newcommand{\Zmin}{Z_{\rm min}}
\newcommand{\zmax}{z_{\rm max}}
\newcommand{\zlight}{z_{\rm light}}
\newcommand{\iP}{{^{\rm intr}P}}
\newcommand{\Puc}{P_{\rm uc}}
\newcommand{\Pcc}{P_{\rm cc}} 
\newcommand{\n}{{\bf n}} 
\newcommand{\N}{{\bf N}} 
\newcommand{\be}{\begin{equation}} 
\newcommand{\ee}{\end{equation}} 
\newcommand{\bi}{\begin{itemize}} 
\newcommand{\ei}{\end{itemize}}

\begin{document}

\leftmargin=-1cm
\title {Fragment size correlations in finite systems - application
to nuclear multifragmentation}
\author{P. D\'esesquelles,\\
IPN, Bt. 100, 15 rue Georges Cl\'emenceau, F91406 - Orsay Cedex, 
France.}

\maketitle

\begin{abstract}

We present a new method for the calculation of fragment size correlations in a
discrete finite system in which correlations explicitly due to the finite
extent of the system are suppressed. To this end, we introduce a combinatorial
model, which describes the fragmentation of a finite system as a sequence of
independent random emissions of fragments. The sequence is accepted when the
sum of the sizes is equal to the total size. The parameters of the model, which
may be used to calculate all partition probabilities, are the intrinsic
probabilities associated with the fragments. Any fragment size correlation
function can be built by calculating the ratio between the partition
probabilities in the data sample (resulting from an experiment or from a Monte
Carlo simulation) and the 'independent emission' model partition probabilities.
This technique is applied to charge correlations introduced by Moretto and
collaborators. It is shown that the percolation and the nuclear statistical
multifragmentaion model ({\sc smm}) are almost independent emission models
whereas the nuclear spinodal decomposition model ({\sc bob}) shows strong
correlations corresponding to the break-up of the hot dilute nucleus into
nearly equal size fragments.\\


\end{abstract}

\section {Introduction}

The break-up of any finite composite system (atomic clusters, atomic nuclei,
fullerenes, molecules\ldots) is characterized by a probability distribution
which incorporates constraints imposed by dynamical or static conservation
laws. Thus, in the case of nuclear decay, the observed multifragmentation modes
provide information on properties of nuclear matter at high excitation energy.
From a statistical view point the simplest fragmentation model may be
formulated by attributing an independent emission probability to each type of
fragment (mass, charge). In the limit of infinite parent system size the
resulting model (which will be referred to herein as the {\it independent
emission model}) exhibits no correlation between fragments. For finite systems,
we show hereafter that the correlations induced by the static conservation
laws, that is in mass and/or charge\footnote{In practice, because of the
difficulties with mass measurements, studies are mainly carried out on charge
partitions, noted $\n:(n_1,\ldots,n_{\ztot})$ where $n_z$ is the number of
charges $z$ in the partition. The charge conservation law reads : $\sum_z z\
n_z =\ztot$.} (referred to hereafter as {\it trivial} correlations), can be
exactly calculated. In the independent emission model, all the physical
information is contained in the emission probabilities of the different types
of particles. However, because of the static conservation laws, these {\it
intrinsic probabilities} are not equal to the observed probabilities. 

Most theoretical multifragmentation models, which describe the process of
instantaneous break-up of the atomic nucleus submitted to extreme temperature
and pressure conditions, introduce other forms of correlations between particle
types. When these correlations are specific to a given model, their
experimental observation constitutes a crucial test of validation/invalidation.
For example, several models, describe the decay of hot nuclei by the
development of density fluctuations (surface or volume instabilities
\cite{Bib_Mor92,Bib_Bae92,Bib_Bor92,Bib_Xu93}). Among these models, the
spinodal nuclear decomposition mechanism may dominate when the collision
between two nuclei leads to a highly heated and sufficiently compressed
nucleonic system. The decompression phase leads the system into the spinodal
zone (zone in which the incompressibility modulus $(\partial P/\partial
\rho)_S$ is negative) where the density fluctuations are exponentially
amplified up to produce fragmentation \cite{Bib_Gua96_Cho95}. The dynamics of
the density waves in the system is dominated by the most un-stable mode, whose
wave length is of the order of 10~fm. Thus the composite nucleus will
disintegrate into almost equal size fragments (in the range $Z \approx 6-20$
and, more particularly, $Z \approx 10-15$, see Section \ref{Sec bob}).

Binary sequential de-excitation models ({\sc gemini}\cite{Bib_Cha88} or {\sc
simon} \cite{Bib_Dur92,Bib_NGu98}) do not exhibit, of course, any preferential
decay into equal charges. Nor do instantaneous multifragmentation models~: the
Copenhagen-Moscow model (code {\sc smm}) \cite{Bib_Bon85} (see Ref.
\cite{Bib_Tab00} for the system Xe+Sn at 32 $A$~MeV, and the Section \ref{Sec
smm} of the present article) and the Berlin model \cite{Bib_Gro90} (code {\sc
mmmc}) \cite{Bib_LeF01}.

Experimentally, the charge distributions are privileged tools for the the study
of nuclear multifragmentation. However, the yields of various charges alone do
not permit a sufficient discrimination of mechanisms. Model validations thus
require the comparison of intra-event charge correlations. In this context, a
difficulty arises from the fact that the detected fragments are not produced
only at the multifragmentation stage of the reaction. Certain light particles,
for example, are emitted during the inter-penetration of the nuclear spheres
(pre-equilibrium phase), others are emitted, at the end of the process, by the
hot multifragmentation fragments. The final (detected) partitions have thus, in
part, lost the memory of the crucial moment of the reaction. Therefore, it may
be necessary to use statistical methods in order to detect the charge
correlations induced by the initial multifragmentation. One of these methods,
proposed by L.~Moretto and collaborators \cite{Bib_Mor96_Mor97}, was shown to
be especially efficient for detecting the presence of the spinodal decay
mechanism (volume instabilities) of the nucleus.

The fragments formed during the spinodal decomposition phase have comparable
sizes (charges). However, this effect is not visible in the charge spectra
generated by a Monte-Carlo code (Brownian One Body Dynamics, {\sc bob} code
\cite{Bib_Gua96_Cho95}) simulating this type of mechanism. The reasons are
numerous~: coalescence and primary fragment deexcitation, finite size effect
inducing mode superpositions \ldots The same remark applies to the experimental
charge distributions~: no excess yield is visible in the expected charge
domain. However, the method of charge correlations reveals, for this model, a
small "fossil" signal which corresponds to events in which the system breaks
into similar size fragments and whose charges have not been modified (or
reduced by the same quantity) before detection. The method introduced by
Moretto and collaborators consists in calculating the correlation function of
the mean charge $\Zmoy$ of the $M$ {\sc imf}\footnote{Intermediate Mass
Fragments. Concretely fragments with charge greater or equal to a given limit
($\Zmin=3$ or 5 in this work).} and of their standard deviation $\sigma$. A
peak appears therefore in this correlation function for $\sigma \approx 0$ and
$\Zmoy \approx10 - 15$. Experimentally, a peak has been effectively observed
for the Xe+Sn system at 32 $A$~MeV in central collisions with the {\sc indra}
multidetector \cite{Bib_Tab00,Bib_Bor01}. Preferential decompositions in three
approximately equal size fragments were also observed in central Xe+Cu reactions
at 45 $A$~MeV with the {\sc multics} multi-detector \cite{Bib_Bru92_Bru94}.

The goals of this article are the following~:

\begin{itemize} 

\item We wish to make the interpretation of correlations more rigorous.
Progress is necessary because the peak related to spinodal decomposition is
often generated by a very small number of experimental or synthesized events.
It corresponds, as we will see, to the ratio of two very small quantities and
therefore will be characterized by a large error bar. The {\it significance}
associated with a peak must therefore be systematically evaluated. To this end,
we show that the error in the denominator of the correlation function can be
greatly reduced by substituting a convolution product for the random selection
process proposed in the initial method.

\item The correlation peak corresponding to the spinodal decomposition (or to
any other cause) is superimposed on a dominant structure due to the
correlations induced by the total charge conservation law (trivial
correlations). This structure often makes the interpretation of the peaks in
terms of physically interesting correlations difficult or ambiguous. Hence, it
is important to correct the correlation function for finite size effects. For
this reason, it has been proposed to construct the denominator of the
correlation function in a different way than that introduced in Ref.
\cite{Bib_Mor96_Mor97} using the minimum information model. It will be shown
that this method can hide peaks corresponding to non trivial correlations.

\item We therefore introduce, in an algebraical exact way, the effects of
charge conservation using the independent emission model. Thanks to this new
method, any event sample with only trivial correlations will show a flat
correlation function.

\item We will study more completely the independent emission model constrained
by the charge conservation. The notion of {\it intrinsic probability} of
particles will be introduced.

\item Finally, this new method will be validated by its application to three
nuclear decomposition models ({\sc smm}, percolation \cite{Bib_Sta85} and {\sc
bob}). It will be shown that these models are, to first order, independent
emission models.

\end{itemize}

\section {The charge correlation function}

\subsection {Algebraic calculation of the denominator}

The quantity $1+R(\Zmoy,\sigma|M)=P(\Zmoy,\sigma|M)/\Puc(\Zmoy,\sigma|M)$,
where $\Zmoy$ is the mean charge of the {\sc imf}, $\sigma$ their standard
deviation and $M$ their multiplicity, will be called the {\it charge
correlation}\footnote{In the following, the variables in capitals will be
relative to the {\sc imf}, the {\sc imf} partitions will be noted $\N$, the
complete partitions $\n$ and the total multiplicity $m$.}. The method
traditionally used to calculate the denominator of a correlation function
\cite{Bib_Mor96_Mor97} consists in constructing "pseudo-events" using randomly
selected fragments belonging to different events of the sample with a given
{\sc imf} multiplicity. The global variable distributions relative to the
pseudo-events do not contain intra-event correlations. The numerator $P$ and
the denominator $\Puc$ of the correlation function are calculated in the same
way, the first one from the sample events, the second one from the
pseudo-events. Since the denominator does not contain intra-event correlations,
its probability density function is written with an index uc (un-correlated).

The only experimental information required for the calculation of the
denominator is the charge distribution of the sample. It is equivalent, and,
from a computational point of view, faster, to sort charges with respect to the
average charge distribution, rather than to select fragments among events. In
fact, the random selection using the charge probability distribution is not
even necessary since the denominator can be calculated algebraically in the form
of a convolution product.

One notes $P(\Zmoy|M)$ the probability to obtain a value $\Zmoy$ of the mean
{\sc imf} charge for the multiplicity $M$ events ($\sum_{\Zmoy} P(\Zmoy|M) =
1$, hereafter all conditional probabilities will be assumed to be normalized by
a relation of the same type). This conditional probability is given by the
convolution~:

\begin{eqnarray}
\Puc(\Zmoy|M) = \sum_{Z_1} \ldots \sum_{Z_{M-1}} P_Z(Z_1|M) \cr
\ldots P_Z(Z_{M-1}|M) \ P_Z(M \Zmoy - (M-1) \Zmoy ' \ |M),
\end{eqnarray}

where $\Zmoy'$ is the mean charge of the {\sc imf} except the last and
$P_Z(Z|M)$ the {\sc imf} charge probability distribution for a given
multiplicity. The last factor accounts for the {\sc imf} total charge
conservation ($\sum_{i=1}^M Z_i = M \ \Zmoy$). The standard deviation is
calculated according to the {\it measure}~:

\begin{eqnarray}
\label{Eq sigma usuel}
\sigma = \sqrt{\frac{1}{M} \ \sum_{i=1}^M (Z_i - \Zmoy)^2}. 
\end{eqnarray}

The equations obtained with the un-biased {\it estimator} of the standard
deviation, used in Ref. \cite{Bib_Mor96_Mor97} are listed in Appendix.
The choice of the expression of the standard deviation will not have any
influence on the shape of the correlation functions nor on the conclusions of
this study. It can be shown that the probability to obtain a standard deviation
$\sigma$, when the fragments are randomly selected, is~:

\begin{eqnarray}
\nonumber
\Puc(\sigma|M) = \sum_{Z_1} \ldots \sum_{Z_{M-1}} P_Z(Z_1|M) \ldots \cr
\times P_Z(Z_{M-1}|M) \ P_Z(\Zmoy ' + M \ 
\sqrt{\frac{\sigma^2}{M-1} - \frac{\sigma'^2}{M}} \ |M)\\ 
\label{Eq p(s|m)}
\times P_Z(\Zmoy ' - M \ 
\sqrt{\frac{\sigma^2}{M-1} - \frac{\sigma'^2}{M}} \ |M) \,,
\end{eqnarray}

where $\sigma'$ is the standard deviation of the charges of the {\sc imf}
except the last. If the term under the square root is negative, the probability
is zero. Finally the correlation between the mean charge and the standard
deviation reads~:

\begin{eqnarray}
\label{Eq P_Z(Z|m)}
\nonumber
\Puc(\Zmoy, \sigma \ |M) = \sum_{Z_1} \ldots \sum_{Z_M} 
P_Z(Z_1|M) \ldots \cr
\times P_Z(Z_{M-1}|M) \ P_Z(Z_M|M)\ \delta_{Z_M , M \Zmoy + (M-1) \Zmoy '} \\ 
\times\delta_{Z_M , \Zmoy ' \pm M \ 
\sqrt{\frac{\sigma^2}{M-1} - \frac{\sigma'^2}{M}}}\ , 
\end{eqnarray}

where $\delta_{a,b}$, the Kronecker symbol, is equal to 1 when $a=b$ and 0
otherwise. The multinomial decomposition leads to an equivalent
(but more practical) form of this equation~:

\begin{eqnarray}
\label{Eq multinomiale}
\Puc(\Zmoy, \sigma \ |M) \ = \cr
\ M! \ 
\sum_{{{\N \atop \sum_Z N_Z = M} \atop \sum_Z Z N_Z 
= M\Zmoy} \atop \sum_Z Z^2 N_Z = M(\Zmoy^2+\sigma^2)} \ 
\prod_Z \frac{P_Z(Z|M)^{N_Z}} {N_Z!} \,,
\end{eqnarray}

where $N_Z$ is the number of {\sc imf} with charge $Z$ and $\N$ an {\sc imf}
partition. The product runs over all possible {\sc imf} charges. The
probabilities in the denominator respect the normalization~: $\sum_{\sigma}
\sum_{\Zmoy} \Puc(\Zmoy, \sigma \ |M)=1$. Hereafter, for notational
simplification, the sum sign of Eq. (\ref{Eq multinomiale}) will be written as
$\sum_{\{\N|M,\Zmoy,\sigma \}}$ and the other sum signs will be formed
according to the same logic.

The extension of this formula of the denominator to samples containing a
variable number of {\sc imf} is useful when the experimental statistics is
reduced. It is expressed straightforwardly as $\Puc(\Zmoy, \sigma) = \sum_M
P_M(M)\,\Puc(\Zmoy, \sigma \ |M)$ (where $P_M$ is the multiplicity probability
distribution of the {\sc imf}), i.e.~:

\begin{eqnarray}
\label{Eq extension M /=}
\Puc(\Zmoy, \sigma)\ =\ \sum_{\{\N|\Zmoy,\sigma\}} P_M(\sum_Z N_Z)\ 
(\sum_Z N_Z)!\ \cr
\prod_Z \frac{P_Z(Z|\sum_{Z'} N_{Z'})^{N_Z}}{N_Z!}\,.
\end{eqnarray}

\subsection {Statistical error bars} 

Let us recall that the correlation function is defined by~:

\begin{equation}
\label{Eq fonc cor}
1+R(\Zmoy, \sigma \ | M) = \frac
{P(\Zmoy, \sigma \ | M)}
{\Puc(\Zmoy, \sigma \ | M)}\,.
\end{equation}

where the probability in the numerator is the number of sample events including
$M$ {\sc imf}, with mean charge $\Zmoy$ and standard deviation $\sigma$,
divided by the number of events with {\sc imf} multiplicity $M$. To first
order, the sampling variance of a proportion applied to Eq. (\ref{Eq fonc cor})
gives the following error~:

\begin{eqnarray}
\label{Eq 1+R(Zmoy,sigma}
\Delta (1+R(\Zmoy, \sigma \ | M))
= \frac
{\sqrt{P(\Zmoy, \sigma \ | M)}}
{\sqrt{N(M)} \ \Puc(\Zmoy, \sigma \ | M)} \,,
\end{eqnarray}

where $N(M)$ is the number of events with $M$ {\sc imf} in the data sample. The
use of formula (\ref{Eq multinomiale}) reduces considerably the statistical
error. In the case of a Monte-Carlo selection process, it would be, to the same
order~:

\begin{eqnarray}
\Delta (1+R(\Zmoy, \sigma \ | M)) =
\frac
{\sqrt{P(\Zmoy, \sigma \ | M)}}
{\sqrt{N(M)} \ \Puc(\Zmoy, \sigma \ | M)} +\cr
\frac
{P(\Zmoy, \sigma \ | M)}
{\sqrt{N_{\rm uc}(M)} \ \Puc(\Zmoy, \sigma \ | M)^{3/2}}\,,
\end{eqnarray}

where $N_{\rm uc}(M)$ is the number of pseudo-events generated by random
selection for the calculation of the denominator. The last term can be very
important in the presence of correlation peaks. The calculation of the error is
crucial when the standard deviation is zero, on the one hand because it is
these events that we are interested in, and, on the other hand, because the
number of events of this type is often very small.

In practical cases, the denominator may be evaluated with a very low
uncertainty thanks to Eq. (\ref{Eq multinomiale}) since only the - very low -
statistical fluctuations on the charge spectrum alter the result. On the other
hand the precision of the numerator depends strongly on the number of events in
the considered sample. Furthermore, the error on the error bar (Eq. (\ref{Eq
1+R(Zmoy,sigma})) depends also on the number of events. Therefore it can be
inaccurate. It would therefore be interesting to obtain an evaluation of the
error bar using only the value of the denominator. This is possible using the
so-called null hypothesis, i.e. that the correlation function is equal to unity
(absence of correlation). In the frame of this hypothesis the error bar is~:

\begin{eqnarray}
\Delta_{1+R=1} (1+R(\Zmoy , \sigma\ | M)) =\cr
\frac
{1}
{\sqrt{N(M)\ \Puc(\Zmoy , \sigma\ | M)}}\,.
\end{eqnarray}

The significance of a positive correlation (of a peak) is defined as being the 
probability, in the frame of the null hypothesis, that the peak has a height
lower than that observed. Therefore, the higher the peak, the higher the
significance. An under-estimation of the significance $\cal S$ can be obtained
straightforwardly using the Schwarz inequality~:

\begin{eqnarray}
{\cal S} \leq
\frac
{N(M) \ \Puc(\Zmoy, \sigma \ | M)^3}
{\left (
P(\Zmoy, \sigma \ | M)-\Puc(\Zmoy, \sigma \ | 
M)\right)^2}.
\end{eqnarray}

Exact calculations of the significance as well as applications to experimental
data will be presented in a forthcoming publication \cite{Bib_Tab01}.

\subsection {Case where all {\sc imf} have the same charge}

\subsubsection {Numerator} 

Since the spinodal decomposition peak is expected when all {\sc imf} have the
same charge, we now consider the case where $\sigma=0$. For a fixed {\sc imf}
mean charge, there is now only one {\sc imf} partition~: $\forall i, Z_i =
\Zmoy$. Thus, differences between the complete partitions with same $\Zmoy$ are
only due to the light fragments whose total charge is $z_{\rm light}=z_{\rm
tot}-M\Zmoy$.

\subsubsection {Denominator}

When $\sigma = 0$, the probabilities given by Eq. (\ref{Eq multinomiale})
become~:

\begin{equation}
\label{Eq P(<Z>,0|M)}
\Puc(\Zmoy , 0\ |M)\ 
\left\{
\begin{array}{l}
=P_Z(\Zmoy \ |M)^M\cr
 \hbox{if }\Zmoy \hbox{ is integer and }\cr
\Zmoy \in [\Zmin,z_{\rm tot}-(M-1)\ \Zmin],\cr
=0 \cr
\hbox{otherwise}\,,
\end{array}
\right.
\end{equation}

where $P_Z(Z|M)$ is the charge distribution for a given {\sc imf} multiplicity
($\sum_{Z=\Zmin}^{\ztot} P(Z|M) = 1$).
The mean charge being equal to the charge of each {\sc imf} implies that
$\Zmoy$ is always an integer. The probability that the standard deviation is
zero is~:

\begin{equation}
\Puc(\sigma=0) \ = \sum_{\Zmoy = \Zmin}^{z_{\rm 
tot}-(M-1)\Zmin} 
P_Z(\Zmoy |M)^M\,.
\end{equation}

When the charge distribution of light {\sc imf} follows a power law or an
exponential law (we will see that this is the case for the minimum information
model), the denominator assumes very simple forms (respectively)~:

\begin{eqnarray} 
\label{Eq tau}
\Puc(\Zmoy , 0\ |M)\ \propto \Zmoy^{-\tau M}\,,\\
\Puc(\Zmoy , 0\ |M)\ \propto {\rm e}^{-\tau M \Zmoy}\,.
\end{eqnarray}

\subsubsection {Correlation function}

As indicated previously, the evaluation of the correlation function in the case
of equal size {\sc imf}, is of considerable physical interest. Unfortunately it
often corresponds to the ratio of two very small probabilities. If the number
of events in the sample is too low, it is possible that no event corresponds to
the given mean charge (the correlation function cannot be calculated) or that a
very small number of events correspond (which can lead to a spurious peak). It
is therefore important to determine, a priori, the minimum number of events
necessary to obtain a reliable evaluation of the correlation function for a
null standard deviation. An evaluation of this number can be obtained making,
once more, the hypothesis that the correlation function is unity. The
probability to obtain an event in which the {\sc imf} charges are all equal to
$\Zmoy$ is $P_Z(\Zmoy|M)^M$ (this quantity can be obtained precisely even with
a reduced event sample). The minimum size $N(M)$ of the sample must be
therefore of one order of magnitude greater than $P_Z(\Zmoy|M)^{-M}$.

\section {Denominator conditioned by charge conservation.}

The formation of the denominator as proposed by Moretto and collaborators 
(that we will continue to call the pseudo-event method though the result is
expressed by the algebraic formula (\ref{Eq multinomiale})) has many
advantages~: it is rigorous, it is simple to evaluate, it takes into account
the efficiency of the detector, it uses only experimentally measured quantities
and the resulting correlation function shows all charge correlations, whatever
their origin. This latter advantage can become an inconvenience when one wishes
to study correlations induced by only one physical cause. In the majority of
cases, the main structure in the correlation function is due to the total
charge conservation law. We will see that this law introduces a large
structure, greater than unity, close to $\Zmoy = z_{\rm tot}/M$. In Ref.
\cite{Bib_Tab00}, this structure was considered as a baseline on which was
superimposed a peak due to the spinodal decomposition mechanism.

In this section we will discuss two different propositions for calculating the
denominator taking into account the charge conservation (in order to remove the
corresponding structure from the correlation function). The first one consists
in using partitions provided by the minimum information model (all the
partitions of a given total charge have the same probability). We will show
that the denominator constructed in this way presents a spurious peak at
$\sigma = 0$ that can conceal a possible physical peak in the correlation
function (Section \ref{Sec info min}). The second proposition consists in
modifying the expression (\ref{Eq multinomiale}) of the algebraic calculation
of the denominator in order to introduce, in an exact way, the influence of
charge conservation with the consequence that charge conservation influences
both numerator end denominator.

\subsection {Minimum information model}

\label{Sec info min}

\subsubsection {Introduction}

In this model, all partitions have the same probability~: $P(\n) = 
1/N(\ztot)$ where $N(\ztot)$ is the total number of partitions for total 
charge $\ztot$. This result is obtained by application of the minimum
information principle (or maximal entropy), information being defined as~:

\begin{equation}
\label{Eq information}
I = \sum_{\n} P(\n) \ln{P(\n)}. 
\end{equation}

Setting the derivative of $I$ equal to zero, under the single constraint of 
charge conservation, one obtains that all probabilities are equal. The total 
number of charge partitions of a charge $\ztot$ nucleus is approximately
given by the Ramanujan-Hardy formula \cite{Bib_Abr65} whose leading term is~:

\begin{equation}
N(\ztot)\approx\frac{\exp{\left(\pi\sqrt{\frac{2\ztot}{3}} \right)}}
{4\ztot\sqrt{3}}\,.
\end{equation}

The number of partitions increases therefore very rapidly with the charge. 
Thus, studies of large systems, by systematic generation of all partitions, are
not possible. The calculation of the number of {\sc imf} partitions (all
fragments have a charge greater than a certain limit) and of light fragments
(fragments with charge lower than a certain limit) is exposed in the companion
article \cite{Bib_Des01}. Some examples of applications of the minimum
information model (possibly modified by combinatorial factors) to nuclear
fragmentation are given in Refs. 
\cite{Bib_Sob85,Bib_Aic84,Bib_Bot00,Bib_DeA89_Mek90_Mek90bis,Bib_Col90_Col94_Col96}

\subsubsection {Case where all {\sc imf} have the same charge}

In this subsection, the correlation function for the minimum information model
in the case $\sigma = 0$ will be calculated exactly. It will be shown that this
function presents a combinatorial peak due to an intrinsic feature of the
model, namely the non-ordering of the charges. We will introduce an alternative
model in which this effect is corrected.

\paragraph {Numerator.}

The numerator $P(\Zmoy,0\ |M)$ of the correlation function is calculated as the
number of partitions with $M$ {\sc imf}, each of charge $\Zmoy$, divided by the
total number of partitions with $M$ {\sc imf}. The charges of the {\sc imf}
being fixed, the number of partitions will be equal to the number of ways to
divide the remaining charge $\zlight=\ztot-M\Zmoy$ into light fragments (i.e.
fragments with charge less than or equal to $\zmax=\Zmin-1$). This constrained
number of partitions will be noted $^{\zmax}N(\zlight)$. Similarly, the number
of partitions of charge $z$ into $M$ fragments with charge greater or equal to
$\Zmin$ will be noted $_{\Zmin}N(z,M)$. These numbers can be calculated exactly
\cite{Bib_Des01}. With our notation, the numerator reads~:

\begin{eqnarray}
\label{Eq num infomin}
P(\Zmoy, 0 \ |M) = \cr
\frac{^{\zmax}N(\zlight)}
{{\sum_{\Zmoy '}} \ {_{\Zmin}N(M\Zmoy',M)} \ ^{\zmax}N(\ztot-M\Zmoy ')}\,.
\end{eqnarray}

\paragraph {Denominator.}

We have seen (Eq. (\ref{Eq P(<Z>,0|M)})) that the denominator is written as
$P_Z(\Zmoy \ |M)^M$ when the standard deviation is zero. The conditional
probability of $\Zmoy$ given $M$ is the number of partitions with $M$ {\sc imf}
weighted by the proportion of charges $\Zmoy$ that they contain, divided by the
number of partitions containing $M$ {\sc imf}, so that~:

\begin{eqnarray}
\label{Eq den infomin}
\Puc(\Zmoy , 0\ |M) =\cr
\left(\frac{\sum_{\Zmoy=\Zmin}^{\ztot/M} \sum_{\{\N|\Zmoy,M\}}
\frac{N_{\Zmoy}}{M}\ ^{\zmax}N(\ztot-M\Zmoy)}
{\sum_{\Zmoy'=\Zmin}^{\ztot/M}\ 
{_{\Zmin}N(M\Zmoy',M)}\ ^{\zmax}N(\ztot-M\Zmoy')}\right)^M
\end{eqnarray}

in which the sum over all partitions containing $M$ {\sc imf} has been
written~:

\begin{eqnarray}
\sum_{\Zmoy=\Zmin}^{\ztot/M}
\sum_{\{\N|\Zmoy,M \}} \equiv 
\sum_{Z_1}
\ldots
\sum_{Z_k=Z_{k-1}} ^ {R_k/(M-k+1)}
\ldots
\sum_{Z_{M}} \,,
\end{eqnarray}

with $Z_0=\Zmin$ and $R_k=\ztot-\sum_{i=1}^{k-1}Z_i$. The charges of the {\sc
imf} are noted $Z_i$ and are written in increasing order.

\paragraph {Correlation function.}

The charge correlation function is thus given by~:

\begin{eqnarray}
\label{Eq 1+R(zmoy,o|M)}
1+R(\Zmoy, 0 \ |M)  = 
^{\zmax}N(\ztot-M\Zmoy)\cr
\times \left({\sum_{\Zmoy '}}
\ {_{\Zmin}N(M\Zmoy',M)} \ ^{\zmax}N(\ztot-M\Zmoy ')\right)^{M-1}\cr
/ \left(\frac{1}{M} \sum_{\Zmoy=\Zmin}^{\ztot/M}\sum_{\{\N|\Zmoy,M \}} 
N_{\Zmoy} \ ^{\zmax}N(\ztot-M\Zmoy)\right)^M
\end{eqnarray}

\begin{figure} 
\epsffile{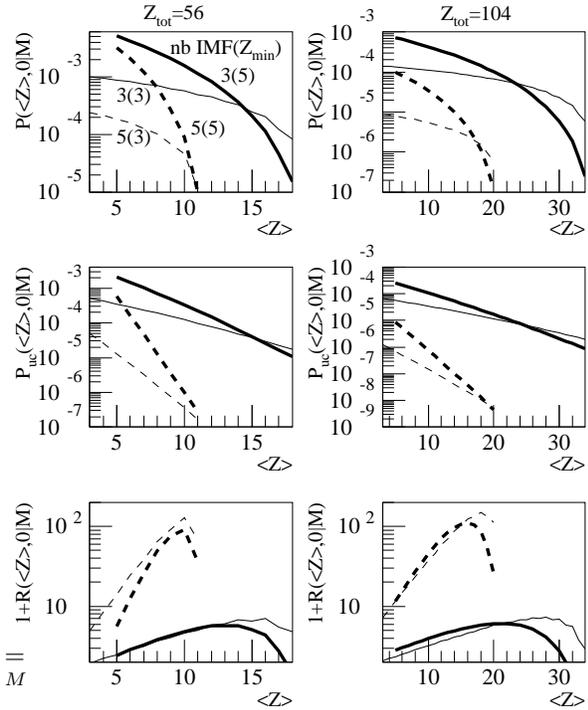} 
\caption { 
\label{Fig numdencor}
Numerator (upper row), denominator (median row) and correlation function 
(lower row) for total charges 56 (left column) and 104 (right column). The bold
lines correspond to {\sc imf} with minimum charge 5, the thin lines to
{\sc imf} with minimum charge 3. The dotted lines correspond to the events
with 5 {\sc imf} and the full lines to those with 3 {\sc imf}. } 
\end{figure}

The latter result is, of course, free of error since it results from the
numbering of all possible partitions. Results for two total charges, two
multiplicities and two definitions of the {\sc imf} are presented in Fig.
\ref{Fig numdencor}.

We observe that~:

\begin{itemize}

\item The behavior of the correlation function depends only weakly on the size
of the system. This result has been observed experimentally~: similar 
correlation functions have been observed for very different systems in central
collisions \cite{Bib_Tab00,Bib_Gui01}.

\item The denominators are exponentially decreasing. This is due to the fact
that the charge distributions are also exponentially decreasing between $\Zmin$
and $\ztot-(M-1)\Zmin$ (Eq. (\ref{Eq tau})).

\item The abscissa of the maximum of the correlation function is a few units 
lower than $\ztot/M$, this property can be used to provide an experimental
determination of the total charge of the composite nucleus, after
pre-equilibrium emission, measuring only the {\sc imf}.

\item The amplitude of the correlation function increases strongly as the
minimum charge of the {\sc imf} diminishes and their multiplicity increases.

\end{itemize}

\begin{figure} \epsffile{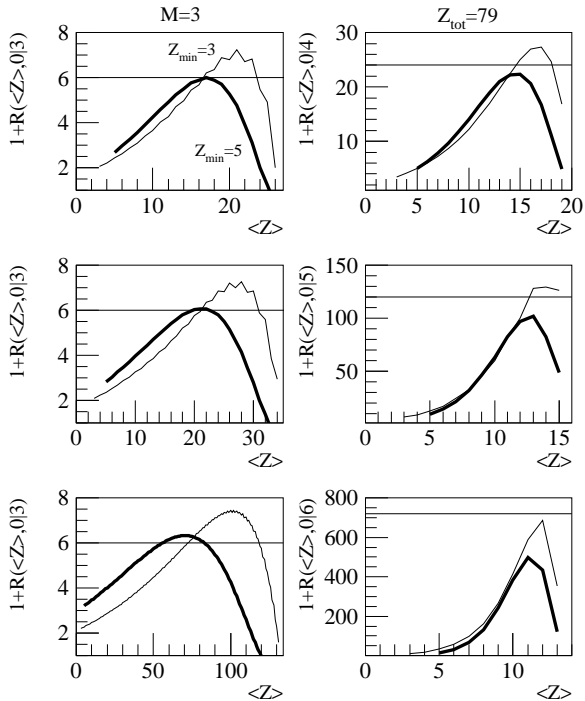} \caption { \label{Fig sig0} Charge
correlation function, for a null standard deviation, obtained by formula
(\ref{Eq 1+R(zmoy,o|M)}) in the framework of the minimum information model.
Heavy traces correspond to {\sc imf} with charges greater than 5 and light
traces to {\sc imf} with charges greater than 3. The correlation functions for
an {\sc imf} multiplicity of 3 are presented on the left hand column in which
the total charges of systems are, from top to bottom, 79, 104 and 400. The
three graphs on the right hand column correspond to the total charge 79 and the
{\sc imf} multiplicity runs from 4 (top) to 6 (bottom). The horizontal lines
are placed at $1+R = M!$.} \end{figure}

Fig. \ref{Fig sig0} presents the correlation function obtained for 
different multiplicity 3 systems (left column) and for the total charge 
79 and different multiplicities (right column).

The left hand column shows that the correlation functions are homothetic for a
given multiplicity. On each figure a horizontal line has been placed at
$1+R(\Zmoy ,0|M) = M!$. One oberves that the maximum of the correlation
function is always of the order of $M!$. This property is due to the
multinomial factors of the denominator (Eq. (\ref{Eq multinomiale})). When the
standard deviation is null, the product $\prod_Z N_Z!$ is identically equal to
$M!$, it factorises therefore in the numerator of the correlation function.
When the standard deviation increases, this product decreases rapidly down to 1
when all charges are different. To illustrate this point, let us consider two
very similar partitions of the total charge 21 into 3 {\sc imf}~: \{7,7,7\} and
\{6,7,8\}. The numerator of the correlation function for both partitions is the
same~: $1/N(21,3) \ (=1/249)$. The denominator is, in the former case,
$P(7|3)^3\ (=6.6\ 10^{-4})$ and, in the latter case, $3!\ P(6|3)P(7|3)P(8|3)\
(=3!\ 6.04\ 10^{-4})$. Therefore the probability product is almost the same in
both cases. Besides, the charge conservation constraint is weak when the {\sc
imf} have small charges hence $1+R(\{6,7,8 \}) \approx 1$ (one finds 1.1).
Consequently $1+R(\{7,7,7 \}) \approx 3!$ (one finds 6.1).

It can be conjectured that this effect would occur for all non-ordered fragment
models. If one multiplies every partition by a factor equal to the number of
charge permutations ($M! / \prod_Z N_Z!$), the peak disappears. On the other
hand probabilities for other sigma values are little modified because the $N_Z$
are then almost always equal to 0 or 1 (i.e. $\prod_Z N_Z!=1$). This latter
model will be referred to as the ordered minimum information model.

\subsection {Algebraic calculation of the denominator with charge conservation}

\subsubsection {Method}

\label{Sec Methode}

The goal of this subsection is to introduce the exact method for the evaluation
of the denominator that eliminates the effects due to charge conservation from
the correlation function. This denominator is obtained by an extension of the 
formula (\ref{Eq multinomiale}) to the whole charge (i.e. including light
fragments). In a first step, we suppose that there is no correlation at all
between charges. This means that each charge is described by a probability
$\iP_z(z)$ (that will be referred to as intrinsic probability of the charge).
The conditional probability of a partition $\n$ (including the {\sc imf} and
the light particles) with total multiplicity $m$ is given then by the
multinomial formula~:

\be
\label{Eq proba part. sans cond.}
P(\n|m) = m!\ \prod_z \frac{\iP_z(z)^{n_z}}{n_z!}\,.
\ee

These conditional probabilities obey the normalization condition $\sum_{\n}
P(\n|m) = 1$. If one introduces the constraint of total charge conservation,
partition constrained conditional probabilities are given by (an index cc will
be applied to probabilities constrained solely by the charge conservation)~:

\be
\label{Eq proba part.}
\Pcc(\n|m) = k(m)\ m!\ \prod_z \frac{\iP_z(z)^{n_z}}{n_z!}\ 
\delta_{\ztot,\sum_z z\,n_z}\ ,
\ee

with~:

\be
\label{Eq k(m)}
k^{-1}(m) = m!
\sum_{\{\n|m,\ztot \}}
\prod_z \frac{\iP_z(z)^{n_z}} {n_z!}\,.
\ee

On the other hand, the multiplicity probability distribution is given by~:

\be
\label{Eq P_m cc}
P_m(m) = \alpha \ m! \,
\sum_{\{\n|m,\ztot \}}
\prod_z \frac{\iP_z(z)^{n_z}} {n_z!}\,,
\ee

with~:

\be
\label{Eq alpha}
\alpha^{-1} = 
\sum_{\{\n|\ztot \}}
(\sum_z n_z)! \,
\prod_z \frac{\iP_z(z)^{n_z}} {n_z!}\,.
\ee

Finally, the partition probabilities are given by~:

\be
\label{Eq Pcc}
\Pcc(\n) = \alpha \ (\sum_z n_z)! \,
\prod_z \frac{\iP_z(z)^{n_z}} {n_z!} \ 
\delta_{\ztot,\sum_z z\,n_z}\ .
\ee

These probabilities contain all the information relative to the charges and to
their correlations. For example, (observed) charge and conditional charge
probability distributions are given respectively by~:

\be
\begin{array}{ll}
\label{Eq Pz(z0),Pz(z0|m)}
P_z(z_0) &= 
\frac{\sum_{\n} n_{z_0}\,\Pcc(\n)}
{\sum_{\n} (\sum_z n_z)\, \Pcc(\n)}\\
&= \frac{\alpha}{\mmoy}\ 
\sum_{\{\n|\ztot\}} n_{z_0}\, (\sum_z n_z)!\,
\prod_z \frac{\iP_z(z)^{n_z}}{n_z!}\,,\\
P_z(z_0|m) &= 
\frac{1}{P_m(m)}
\sum_{\{\n|m\}} \frac{n_{z_0}}{m}\ \Pcc(\n)\\
&= \frac{\sum_{\{\n|m,\ztot\}}
n_{z_0}\,\prod_z \frac{\iP_z(z)^{n_z}}{n_z!}}
{m\,\sum_{\{\n|m,\ztot\}}
\prod_z \frac{\iP_z(z^{n_z})}{n_z!}}\,.
\end{array}
\ee

Because of correlations, the intrinsic probabilities are not equal to the
probabilities to observe the charges (the conservation constraint favours small
charges, see Fig. \ref{Fig tpc mec}b). Equality between intrinsic probabilities
and observed probabilities is valid only for an infinite system.

The intrinsic probabilities are quantities which are not directly measurable,
they must be calculated by inversion of Eq. (\ref{Eq Pcc}) where the $\Pcc(\n)$
are the measured frequencies of the partitions. The set of Eqs. (\ref{Eq
Pz(z0),Pz(z0|m)}) constitutes an under-determined system. It is thus not
possible to obtain an unique solution. 

However, inversion of (\ref{Eq Pcc}) is possible if the non trivial
correlations between charges in the studied sample are weak. When the intrinsic
probabilities are determined, the probabilities of the denominator can be
calculated by summing the complete partition probabilities having the same {\sc
imf} mean charge and standard deviation~:

\be
\Pcc(\Zmoy, \sigma | M) = \frac{1}{P_M(M)} 
\sum_{\{\n|M,\Zmoy,\sigma \}}
\Pcc(\n) \,.
\ee

Finally we can write that the denominator is constructed using the intrinsic
probabilities~:

\begin{eqnarray}
\label{Eq deno ztot}
\Pcc(\Zmoy , \sigma | M) =\cr 
\frac
{\sum_{\{\n|\ztot,M,\Zmoy,\sigma\}}\,(\sum_z n_z)!\,
\prod_z \frac{\iP_z(z)^{n_z}}{n_z!}}
{\sum_{\{\n|\ztot,M\}}\,(\sum_z n_z)!\,
\prod_z \frac{\iP_z(z)^{n_z}}{n_z!}}\,.
\end{eqnarray}

This new denominator takes explicitly into account the charge conservation. 
Structures observed in the corresponding correlation function will
necessarily arose from other causes.

\subsubsection {Monte-Carlo generation of events without non trivial charge
correlations}

A sample of events without non trivial correlations may be synthesized by the
following procedure. Charges $z_i$ are selected randomly according to their
intrinsic probabilities until $\sum_i z_i \geq \ztot$. The event is preserved
only if $\sum_i z_i=\ztot$. The resulting charge spectra are those given by
Eqs. (\ref{Eq Pz(z0),Pz(z0|m)}). The alternative procedure that would
consist in randomly selecting $M-1$ charges, and deducing the last charge using
charge conservation would introduce a bias since the distribution of the last
charge would be different from the preceding ones.

\subsubsection {Combinatorial independent emission models}

We stated previously that models can be characterized by the term "independent
emission" if they contain no correlations, other than those induced by the
conservation of the total charge. This definition implies that partition
probabilities generated by such models can be written in the form of Eq.
(\ref{Eq Pcc}). We give here two examples of independent emission models.

\paragraph {Ordered minimum information model.}

\begin{figure}
\epsffile{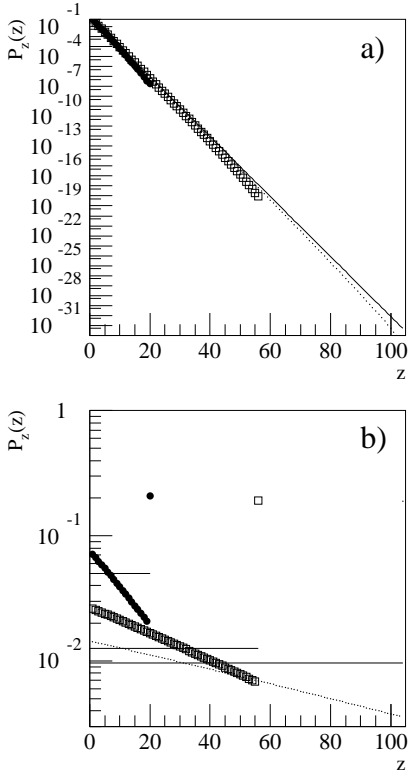}
\caption {
\label{Fig tpc mec}
Charge distributions obtained in the ordered minimum information model (a) and
in the charge equi-probability model (b) constrained by total charge
conservation. Circles correspond to a total charge of 20, squares to
$\ztot=56$ and points to $\ztot=104$. The full lines are the intrinsic
probability distributions.}
\end{figure}

The model produced by weighting partitions by a factor $m!/\prod_z n_z!$ is an
independent emission model. One notes that this weighting is the same as the
one of Eq. (\ref{Eq Pcc}) if the product $\prod_z {\iP_z}(z)^{n_z}$ is
constant. This condition is fulfilled if and only if $\iP_z(z)=a^z$ (the
intrinsic probability product is $a^{\ztot}$). Thus, the normalization of the
probabilities is $\sum_z a^z = 1$. For a sufficiently large total charge (in
practice, superior to 10), the normalization condition implies $a=1/2$. The
model in which every partition is weighted by the number of permutations of its
charges ($m!/\prod_z n_z!$) is therefore an independent emission model with
intrinsic probabilities $\iP_z(z)=2^{-z}$. The resulting charge distributions
are almost exponentially decreasing (Fig. \ref{Fig tpc mec}a). Conversely, the
minimum information model is not an independent emission model (it is not
possible to find a set of intrinsic probabilities such that all partition
probabilities would be the same).

\paragraph {Model of charge equi-probability.}

The minimum information model is taken as implying the equi-probability of the
partitions. In this context it is also interesting to see results given by
another elementary model obtained by assuming intrinsic equi-probability of
the charges. The resulting charge distribution is of course not uniform
because of the constraint of conservation of the total charge.

The Eq. (\ref{Eq Pcc}), for $\iP_z(z) =1/\ztot$, gives~:

\be
\begin{array}{lll}
\label{Eq Pcc MEC}
\Pcc(\n) 
&=& \alpha\,(\sum_z n_z)!\,\prod_z 
\frac{\left(\frac{1}{\ztot}\right)^{n_z}}{n_z!}\ 
\delta_{\ztot,\sum_z z\,n_z}\\
&=& \frac{\alpha\,m!}{\ztot^m\,\prod_z n_z!}\ 
\delta_{\ztot,\sum_z z\,n_z}\ ,
\end{array}
\ee

In other words, if one generates partitions by imposing charge conservation and
if all charges have the same intrinsic probability to be selectioned, then the
partition weight is $m!/(\ztot^m\,\prod_z n_z!)$. Examples of observed charge
spectra are given for different values of the total charge in Fig. \ref{Fig tpc
mec}b). These spectra are very different from that of the intrinsic
probabilities. Indeed one observes an almost exponential decrease. Only the
last charge has a probability which is not consistent with this tendency. This
behaviour is due to the fact that the charge $\ztot-1$ can appear accompanied
only by a charge 1, whereas the charge $\ztot$ is never rejected. More
precisely, Eq. (\ref{Eq Pz(z0),Pz(z0|m)}) gives $P_z(\ztot) = (\ztot/2) \
P_z(\ztot-1)$. One notices that the greater the total charge the less the
influence of the constraint of conservation and consequently the smaller the
slope of the exponential.

\subsubsection {Multiplicity constrained independent emission model}

We now consider the model for which probabilities of partitions with fixed 
multiplicity are given by Eq. (\ref{Eq proba part. sans cond.}), but the
multiplicity probability distribution is not given by the combinatorial Eq.
(\ref{Eq P_m cc}). The multiplicity probability distribution $P_m(m)$ is
imposed a priori. This model has been studied extensively by A.J.~Cole and
collaborators \cite{Bib_Col90_Col94_Col96}. The main difference between the
quoted articles and the study presented in this paragraph resides in the
interpretation of quantities noted $X_z$ by Cole {\it et al.} and, here,
$\iP_z$ as, respectively, adjustable parameters and intrinsic probabilities.
However, the $X_z$ parameters being defined to a factor $a^z$, it is always
possible to normalize them. Indeed, the product $\prod_z (a^z X_z)^{n_z}$ is
equal to the product $\prod_z X_z^{n_z}$ to within a constant ($a^{\ztot}$).
Eq. (\ref{Eq proba part. sans cond.}) and the normalizations give~:

\begin{eqnarray}
\label{Eq proba part. mult contr}
\Pcc(\n) = \frac{P_m(m)}
{\sum_{\{\n'|m,\ztot\}}\prod_z \frac{\iP_z(z)^{n'_z}}{n'_z!}} \cr
\times \prod_z \frac{\iP_z(z)^{n_z}}{n_z!}\ 
\delta_{\ztot,\sum_z z\,n_z}\ ,
\end{eqnarray}

with~:

\be
m = \sum_z n_z\,.
\ee

In the same way as for charges, one can introduce an intrinsic probability for 
multiplicities ($\iP_m$). The partition probability is then written as~:

\be
\Pcc(\n) = k \ \iP_m(m) \ m! \ 
\prod_z \frac{\iP_z(z)^{n_z}} {n_z!} \ 
\delta_{\ztot,\sum_z z\,n_z}\ ,
\ee

with~:

\be
k^{-1} = \sum_m\ \iP_m(m)\ m!
\sum_{\{\n|m,\ztot\}}
\prod_z \frac{\iP_z(z)^{n_z}}{n_z!}\,.
\ee

One deduces the relation between observed probabilities of multiplicities and
those of the intrinsic probabilities~:

\begin{eqnarray}
P_m(m) =\ \iP_m(m)\cr
\left[
\frac{m!\ \sum_{\{\n|m,\ztot\}}\ \prod_z \frac{\iP_z(z)^{n_z}}{n_z!}}
{\sum_{m'}\ \iP_m(m')\ m'!\ \sum_{\{\n|m',\ztot\}}
\prod_z \frac{\iP_z(z)^{n_z}}{n_z!}}
\right]
\end{eqnarray}

Intrinsic probabilities and observed probabilities possess the same 
distributions in the limit of an infinite system size.

\section {Applications}

\subsection {Introduction}

In this section we show the charge correlation obtained, for several nuclear
decay models, using the denominator given by the independent emission
hypothesis. The first step of the procedure consists in determining the
intrinsic probabilities of the charges for each model sample. These
probabilities are obtained by a recursive procedure of minimization of the
$\chi^2$ between probabilities of partitions in the synthesized sample and
those given by Eq. (\ref{Eq proba part.}). The convergence of the procedure is
possible only if non trivial correlations between charges are weak. The minimum
$\chi^2$ is therefore an indication of the strength of these correlations. The
second step, calculation of the denominator by the method presented in Section
\ref{Sec Methode} (\ref{Eq deno ztot}), would also not be possible in the
presence of strong correlations. We will see that this condition is fulfilled
by the three models that we are going to study. Results of the application of
this procedure to the experimental events will be presented in forthcoming
articles \cite{Bib_Tab01,Bib_Gui01}.

\subsection {The Copenhagen model}

\label{Sec smm}

The Copenhagen model \cite{Bib_Bon85,Bib_Bon85bis_Bon95} is a hot liquid drop
model that describes the multifragmentation of the nucleus as an instantaneous
statistical mechanism. The probability of a partition in mass and in charge
$n_{a,z}$ for an excitation energy $E^*$ is given by~:

\begin{eqnarray}
\label{Eq smm}
P(\n |E^*) = P_I(E^*)\ 
\left[
\frac{V_F}{\lambda_T^3}
\right]^m 
\left[
\prod_{a,z} \frac{a^{3n_{a,z}/2}}{n_{a,z}!}
\right]\cr
\times \delta_{a_{\rm tot} , \sum_{a,z} a\ n_{a,z}}\ 
\delta_{\ztot , \sum_{a,z} z\ n_{a,z}}\ .
\end{eqnarray}

The two Kronecker symbols account for the conservation of mass and of charge.
The de Broglie wavelength of the nucleon $\lambda_T=\sqrt{2\pi\hbar/(m_n T)}$
depends on the temperature which is roughly constant at fixed excitation
energy. The density of states corresponding to the internal excitation energy
of the fragments $P_I(E^*)$ is also constant at a given excitation energy. In
this model, as in comparable models, the multiplicity is correlated linearly
with the excitation energy, a prediction which is verified by experimental
observation. The first two factors can be considered therefore to be constant
for a given multiplicity. The volume of the nucleus at the time of
fragmentation, $V_F$, is supposed to be independent of the partition (or,
according to the version of the model, dependent only on the multiplicity). If
we disregard the conservation of the mass, the emission probability of a
fragment is proportional to $a^{3/2}$~: $P_z(z) = \alpha \, a^{3/2}$, so that
the product $\prod_z P_z(z)^{n_z} = \alpha^m \prod_z a^{3n_z/2}$ involves a new
factor depending only on the multiplicity. The equation can therefore be
re-written as~: 

\begin{equation}
P(\n |m) \propto \prod_z \frac{P_z(z)^{n_z}} {n_z!} \ 
\delta_{z_{\ztot} , \sum_z z \ n_z}\ ,
\end{equation}

so that we recover the same expression as that obtained for the independent
emission model. We thus expect that the correlation function is everywhere
equal to 1 and if the denominator is calculated from pseudo-events, we expect
that the shape of the correlation function is determined by charge
conservation. However, this conclusion is based on a simplified form of the
model. The correlation function may thus exhibit weak modulations. Moreover,
one does not expect a peak for small values of the standard deviation.

\begin{figure}[htb]
\epsffile{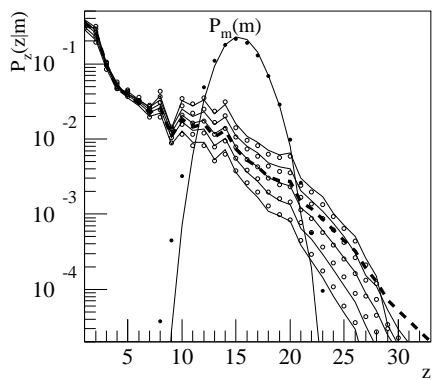}
\caption {
\label{Fig smm zm}
Comparison of charge spectra for several multiplicities for {\sc smm} (circles)
and for the independent emission hypothesis (lines). The dashed line
corresponds to the intrinsic probabilities. The multiplicity probability
distributions, $P_m(m)$, given by the two models are also shown in the figure.}
\end{figure}

Using the {\sc smm} code, 35 million events have been generated for the
$^{138}$Ba nucleus excited to 5 MeV per nucleon. We built, from this sample,
the charge correlation function for the {\sc imf}. The Copenhagen model
produces results (almost) consistent with independent emission as shown in Fig.
\ref{Fig smm zm}. Discrepancies can be explained notably by the fact that the
hot fragments produced during the multifragmentation phase described by Eq.
(\ref{Eq smm}), thereafter decay by light particle emission.

\begin{figure}[htb]
\epsffile{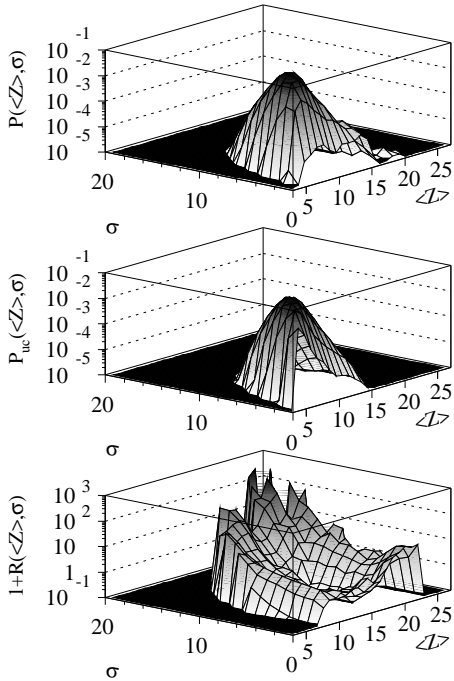}
\caption {
\label{Fig Mor smm}
Upper and lower figures present, in order, the numerator, denominator and 
$[\Zmoy,\sigma]$ correlation function for the Copenhagen model. The denominator
is calculated by the method of pseudo-events (formula (\ref{Eq
multinomiale})).}
\end{figure}

\begin{figure}
\epsffile{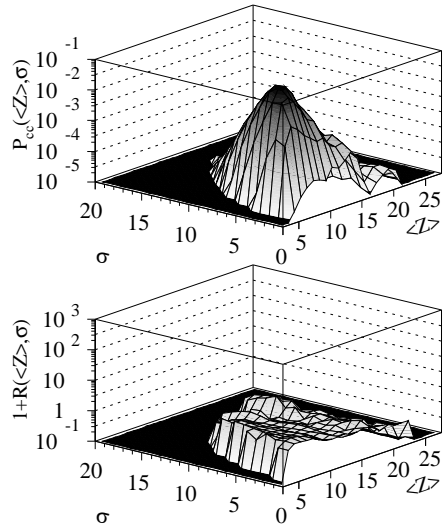}
\caption {
\label{Fig Des smm}
Denominator (upper figure) and correlation function (lower figure) of 
$[\Zmoy,\sigma]$ for the Copenhagen model (the numerator is presented at the 
top of the previous figure). The denominator is calculated using the 
independent emission hypothesis, via formula (\ref{Eq deno ztot}).} 
\end{figure}

The charge correlation functions calculated with the denominators of formulas 
(\ref{Eq multinomiale}) and (\ref{Eq deno ztot}) are presented in the Figs.
\ref{Fig Mor smm} and \ref{Fig Des smm}. In the first case, the main structures
are due to the conservation of the charge, in the second, the correlation
function is practically flat. Non trivial correlations between charges are
therefore very weak for this model and no preferential fragmentation into equal
charge is observed. The - small - modulations of the correlation function are
due to physical causes and to statistical fluctuations.

\subsection {Percolation}

\begin{figure}
\epsffile{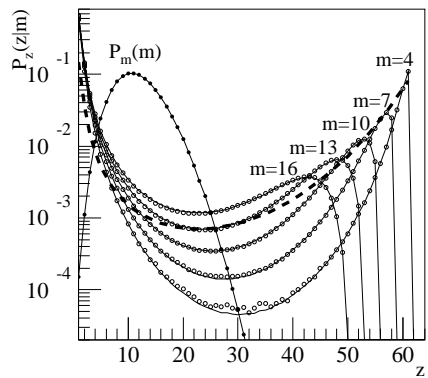}
\caption {
\label{Fig per zm}
Percolation (same conventions as in Fig. \ref{Fig smm zm}.)} 
\end{figure}

The same study has been carried out on a sample obtained with a percolation
code \cite{Bib_Sta85}. A sample of 10$^8$ events has been generated using a 3D
percolation program on a simple cubic 4$\times$4$\times$4 periodic frame, for a
bond breaking probability of 70 \%. The result of the $\chi^2$ minimization
process is presented Fig. \ref{Fig per zm} in which are compared charge
distributions for various multiplicities in percolation and those given by the
formula (\ref{Eq proba part.}) (the intrinsic probabilities are indicated by a
dotted line). The very good agreement indicates that the correlations between
charges are very weak in this model. 

\begin{figure}[htb]
\epsffile{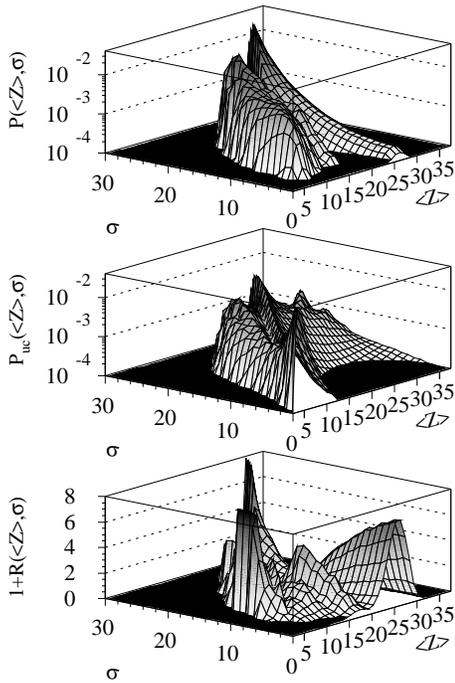}
\caption {
\label{Fig Mor per}
Percolation (same conventions as in Fig. \ref{Fig Mor smm}.)} 
\end{figure}

\begin{figure}
\epsffile{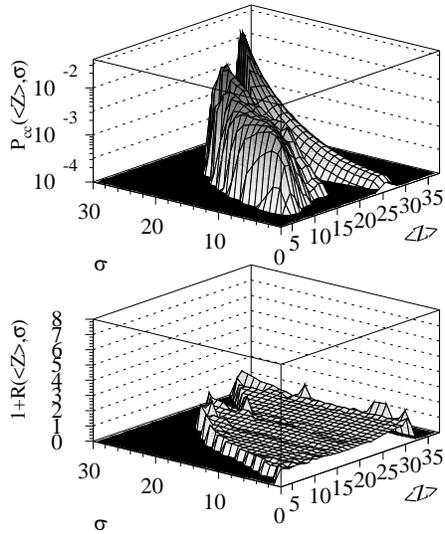}
\caption {
\label{Fig Des per}
Percolation (same conventions as in Fig. \ref{Fig Des smm}.)} 
\end{figure}

Figs. \ref{Fig Mor per} and \ref{Fig Des per} present percolation correlation
functions obtained using denominators given respectively by the pseudo-event
method (Eq. (\ref{Eq multinomiale})) and by the independent emission hypothesis
(Eq. (\ref{Eq deno ztot})). These figures were constructed for all possible
values of the number of {\sc imf}. In the first case, structures are almost
entirely due to the conservation of the charge (to each {\sc imf} multiplicity
corresponds an edge line). When the denominator is calculated using the
intrinsic probabilities, the correlation function is flat and equal to 1 (Fig.
\ref{Fig Des per}). The small peaks on the sides of the correlation function
are due to the statistical fluctuations.

The model of percolation can therefore be assimilated to knowledge of the
intrinsic probabilities.

\begin{figure}
\epsffile{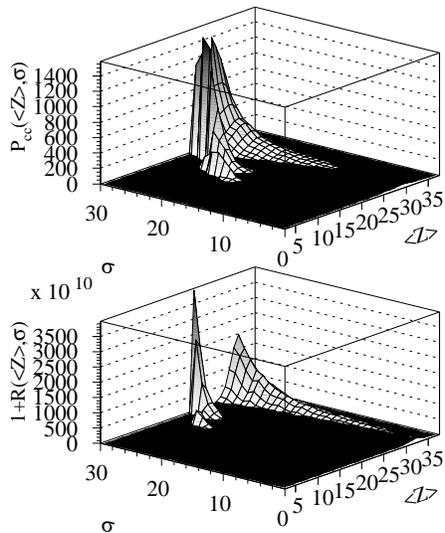}
\caption {
\label{Fig TPP per}
Correlation function of $[\Zmoy,\sigma]$ for the percolation process (the
numerator is presented at the top of Fig. \ref{Fig Mor per}). Denominators are
given by the models of the minimum information (upper figure) and of the
ordered minimum information (lower figure).}
\end{figure}

Fig. \ref{Fig TPP per} presents the correlation function of the percolation
calculation when the denominator is calculated with the minimum information
model (upper figure) and with the ordered minimum information model (lower
figure) for same total charge ($\ztot = 64$). In both cases, the correlation
function presents large structures which are not easily interpreted.

\subsection {The Brownian One Body dynamical model}

\label{Sec bob}

\begin{figure}[htb]
\epsffile{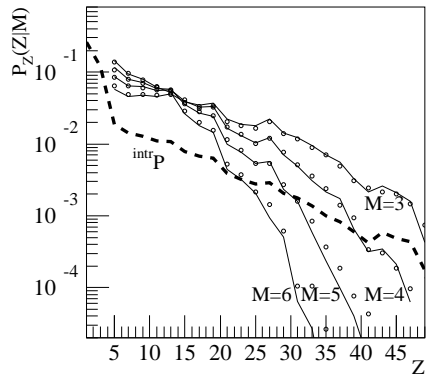}
\caption {
\label{Fig bob zm}
Model Brownian One Body (same conventions as in Fig. \ref{Fig smm zm}.)}
\end{figure}

The last model to be studied in this work is characterized by non trivial
correlations, even though, as motioned in the introduction, they are partly
masked by different processes. The preferential decomposition of the system
into almost equal charges (in the range $[10,20]$), which characterizes the
{\sc bob} model, is not visible, for example, in the inclusive charge spectra
(Fig. \ref{Fig bob zm}).

Our simulated sample was obtained via a four step process \cite{Bib_Fra98}. The
collision entrance channel has been simulated using a one-body semi-classical
microscopic calculation of the {\sc bnv} type \cite{Bib_Bon94} for the system
$^{129}$Xe + $^{119}$Sn at 32 $A$ MeV. This calculation shows that, for the
most central collisions, a compressed single source is formed, after a weak
pre-equilibrium emission, within 40 fm/c. The decompression phase, the entrance
into the spinodal zone and the formation of the fragments are then followed by
the {\sc bob} code which simulates Boltzmann-Langevin density fluctuations
\cite{Bib_Bix69} and the evolution of the system density submitted to spinodal
instabilities. In a third step, fragments are formed using an algorithm which
regroups contiguous cells which numbers of test particles (40 test particles
are used to simulate a nucleon) is greater than a given threshold. The
resulting nuclei are hot, their statistical decay and their Coulomb expansion
are, in a last step, simulated by the {\sc simon} code
\cite{Bib_Dur92,Bib_NGu98}. The event samples are eventually filtered using the
{\sc indra} response function \cite{Bib_Cus}.

The regrouping of pseudo-particles generated by the model is possible only for 
fragments of charge superior or equal to 5. The light fragments are not known. 
This difficulty must be taken into account in the routine of intrinsic
probability optimization~: the probability of a partition of $M$ {\sc imf} is
the sum of probabilities of all partitions containing these $M$ {\sc imf}
(solely) together with the corresponding light particles~:

\be
P(\N) = \alpha\ \sum_{\{\n|\ztot,\N\}} (\sum_z n_z)!\ 
\prod_z \frac{\iP_z(z)^{n_z}}{n_z!}
\ee

The calculation of these probabilities can be accelerated considerably by 
noticing that they can be written in the form~:

\begin{eqnarray}
P(\N) = \alpha \ \left(M! \ \prod_Z \frac{\iP_Z(Z)^{N_Z}} {N_Z!} \right) \cr
\times \left(\sum_m {\rm C}_m^{M+m} \ k(m,\ztot-\sum_Z Z N_Z)\right)
\end{eqnarray}

The last factor depends only on the sum of charges of the {\sc imf} and on
their multiplicity ($k$ is given by Eq. (\ref{Eq k(m)})). The result 
of the fit of the intrinsic probabilities is given in Fig. \ref{Fig bob 
zm}.

\begin{figure}
\epsffile{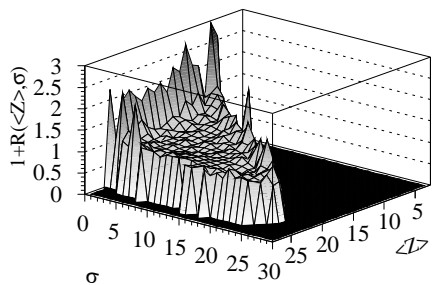}
\caption {
\label{Fig Des bob}
Brownian One Body model correlation function (same conventions as in Fig.
\ref{Fig Des smm}.)}
\end{figure}

One notices that, in spite of the absence of light particle in the sample, the 
intrinsic probabilities have a realistic distribution for all charges. The 
resulting correlation function is presented in Fig. \ref{Fig Des bob}. The
partition probabilities are correctly reproduced by the independent emission
hypothesis, hence the correlation function is practically flat. However, in
contrast with the previous models, it includes strong correlation peaks near
$\sigma = 0$, and, to a lesser extent, for the maximal values of $\sigma$ for
given $\Zmoy$. These latter peaks (as well as those corresponding to $\sigma
=0$, $\Zmoy\leq 9$) have a low significance, so they can only be due to the
statistical fluctuations. Peaks at $\sigma =0$, $\Zmoy \geq 10$, on the other
hand, are meaningful. They signify the spinodal decomposition of nucleus
produced by the {\sc bob} code. 

\section {Conclusions}

The main goal of this article was to present a new method for the evaluation of
the non trivial correlations between the fragment sizes of a finite size
system. The conclusions are the following~:

\bi

\item The Monte-Carlo calculation of the denominator proposed by Moretto and
collaborators can be replaced by a fast algebraic calculation which is
equivalent to the selection of an infinite number of pseudo-events (\ref{Eq
multinomiale}). This calculation results in decreased error bars in the
correlation function notably those associated with correlation peaks (\ref{Eq
1+R(Zmoy,sigma}). The extension of the calculation of the correlation function
to samples including a variable number of {\sc imf} presents no particular
difficulty (\ref{Eq extension M /=}).

\item Correlation functions thus obtained for the different models studied in
this article ({\sc smm}, percolation, minimum information, {\sc bob}) all
possess one maximum for $\Zmoy$ smaller by a few units than $\ztot/M$. This
property leads to the evaluation of the size of the composite nuclei for which
decays have been observed experimentally.

\item This type of denominator possesses many advantages. However, 
correlations induced by charge conservation are always important. They may
conceal other less trivial correlations, or distort the evaluation of their
amplitude. It would, therefore, be useful to define a conventional method for
calculation of the denominator including effects induced by the charge
conservation.

\item It has been proposed to evaluate the denominator using the minimum
information model (all possible partitions of a given total charge has the same
probability). This model incorporates charge conservation but possesses a
purely combinatorial correlation peak at $\sigma = 0$ so that there is a risk
of concealing a physical peak present in the data sample (this effect can be
corrected by the weighting of the partition probability by the number of its
permutations~: $m!/\prod_z n_z$). Furthermore, the numerator (physical sample)
and denominator (given by this model) may correspond to distinct charge and
multiplicity distributions. Finally, the correlation function presents numerous
structures which are difficult to interpret.

\item The two previous conclusions lead us to propose a new calculation of the
denominator, the goal being to replicate all features of partitions of the
numerator excluding intra-event correlations due to other reasons than charge
conservation. In the case where these non trivial correlations are weak, this
goal is reached exactly using the independent emission hypothesis constrained
by the conservation of the charge. Probabilities of partitions are given by the
formula (\ref{Eq Pcc}) which is based on the specification of intrinsic
probabilities for each charge. These values represent probabilities for a
charge to be observed if the constraint of charge conservation played no role.
The intrinsic probabilities are not observables, so that they must be searched
for by a procedure of minimization between probabilities of partitions in the
data sample and those given by the formula (\ref{Eq Pcc}). If the resulting
$\chi^2$ is low, it means that the studied sample is essentially composed of
events corresponding to independent emission. The partition correlation
function (i.e. the set of ratios of the probabilities of the sample divided by
the probabilities given by Eq. (\ref{Eq Pcc})) must then be always near unity
except possibly for a reduced number of partitions corresponding to the non
trivial correlations. In this work, only the $[\Zmoy,\sigma]$ correlation has
been studied, but the same procedure can apply to any type of correlation.

\item The proposed method has been applied to three models of nuclear
multifragmentation. It has been shown that all three models correspond to
almost independent emission. The first two (percolation and the {\sc smm}
statistical multifragmentation code) result in correlation functions everywhere
equal to 1 (to within 10 \%). The code {\sc bob}, on the other hand, exhibits a
flat correlation function everywhere except at $\sigma = 0$. These correlation
peaks are due to the mechanism of spinodal decomposition that favours
partitions which include {\sc imf} of the same charge. These results legitimate
the use of the charge correlation functions method for the experimental search
for spinodal decomposition.

\ei

In forthcoming articles we will study problems related to the application of
this method to experimental event samples (superposition of sources ,
distribution of total charge, pre-equilibrium emission, experimental
efficiency, calculation of the significance of the result) and we will present
results obtained by the {\sc indra} collaboration for heavy ion central
collisions near the Fermi energy.

I wish to thank I.N.~Mishustin who suggested this work, M.F.~Rivet, B.~Borderie
and M.~Parlog for numerous and fruitful discussions and A.J.~Cole for precious
advice concerning the manuscript.

\appendix

\section*{Equations resulting from the utilization of the non-biaised
estimator of the standard deviation}

In this article, we used the usual definition of the standard deviation
(\ref{Eq sigma usuel}). The authors of Ref. \cite{Bib_Mor96_Mor97} preferred to
use the non-biased estimator (in this sense that the mean of its sampling
function is equal to the real value). In writing this article, $\sigma$ being
used as a measure and not as an evaluation of the standard deviation of an
unknown distribution we restricted ourselves to its usual definition. In this
Appendix we give the equations which result from the use of the non-biased
value, $\sigma_{\rm nb}$~:

\begin{eqnarray}
(\ref{Eq sigma usuel}) & \rightarrow 
&\sigma_{\rm nb} = \sqrt{\frac{1}{M-1}\ \sum_{i=1}^M (Z_i - \Zmoy)^2}\\
\nonumber
(\ref{Eq p(s|m)}) & \rightarrow &\Puc(\sigma_{\rm nb}|M) = 
\sum_{Z_1} \ldots \sum_{Z_{M-1}}\\
\nonumber
&&\times P_Z(Z_1|M) \ldots P_Z(Z_{M-1}|M)\\ 
\nonumber
&&\times P_Z(\Zmoy'+\sqrt{M\sigma_{\rm nb}^2-
\frac{M(M-2)}{M-1}\sigma_{\rm nb}'^2}\ |M)\\ 
\nonumber
&&\times P_Z(\Zmoy'-\sqrt{M\sigma_{\rm nb}^2-
\frac{M(M-2)}{M-1}\sigma_{\rm nb}'^2}\ |M)\\
(\ref{Eq P_Z(Z|m)}) & \rightarrow 
&\Puc(\Zmoy , \sigma_{\rm nb}\ |M) = \sum_{Z_1} \ldots 
\sum_{Z_M} \\
\nonumber
&&\times P_Z(Z_1|M) \ldots P_Z(Z_{M-1}|M)\ P_Z(Z_M|M)\\
\nonumber
&&\times \delta_{Z_M , M \Zmoy + (M-1) \Zmoy'}\\
&&\times 
\delta_{Z_M , \Zmoy'\pm 
\sqrt{M\sigma_{\rm nb}^2 - \frac{M(M-2)}{M-1}\sigma_{\rm nb}'^2}}\\ 
\nonumber
(\ref{Eq multinomiale}) & \rightarrow 
&\Puc(\Zmoy , \sigma_{\rm nb}\ |M)\ =\ M!\\
&& \times \sum_{{{\N\atop 
\sum_Z N_Z = M} \atop 
\sum_Z Z N_Z = M\Zmoy} \atop 
\sum_Z Z^2 N_Z = M \Zmoy ^2+(M-1)\sigma_{\rm nb}^2}\ 
\prod_Z \frac{P_Z(Z|M)^{N_Z}}{N_Z!}
\end{eqnarray}

The other equations are not modified.

\end{document}